\documentclass[12pt,preprint]{aastex}

\usepackage{emulateapj5}

\shorttitle{Comments on ``EBL detections'' by Bernstein et al.}
\shortauthors{Mattila}

\newcommand{\cgs}{$10^{-9}$erg\,s$^{-1}$cm$^{-2}$sr$^{-1}$\AA$^{-1}$\,}

\begin{document}

\title{COMMENTS ON  ``THE FIRST DETECTIONS OF THE EXTRAGALACTIC BACKGROUND\\
 LIGHT AT 3000, 5500, AND 8000 \AA'' BY BERNSTEIN, FREEDMAN AND MADORE}

\author{K. Mattila\altaffilmark{}}
\affil{Observatory, University of Helsinki, P.O. Box 14, FIN-00014 Helsinki, Finland}
\email{mattila@cc.helsinki.fi}

\begin{abstract}
A critical discussion is presented of the data analysis
applied by  \citet[]{bfm02a, bfm02b} in their measurement of the Extragalactic 
Background Light. 
There are questionable assumptions in the analysis of the ground-based
observations of the Zodiacal Light. The modeling of the Diffuse Galactic
Light is based on an underestimated value of the dust column density along the 
line of sight. Comparison with the previously presented results from the same 
observations reveals a puzzling  situation: in spite of a large 
difference in the atmospheric scattered light corrections
the derived  Extragalactic Background Light values are exactly the same.   
The claim of the paper of a ``detection of the Extragalactic Background Light'' 
appears premature. 
\end{abstract}


\section{Introduction}

\citet[]{bfm02a, bfm02b}, hereafter BFM02a, BFM02b, or BFM02 combined,  
have announced the first detection of the Extragalactic Background Light 
(EBL) from absolute photometry, with the mean values of
4.0($\pm2.5$), 2.7($\pm1.4$), and 2.2($\pm1.0$) $\times$\cgs\, at
3000, 5500, and 8000 \AA\,, respectively. The errors quoted are
1$\sigma$ uncertainties.
Their method  is based on the formula:
\begin{equation}
I_{\rm EBL} = I_{\rm tot} - I_{\rm ZL} - I_{\rm DGL},
\end{equation}
where
$I_{\rm tot}$ is the total sky surface brightness outside the atmosphere,  
$I_{\rm ZL}$  is the Zodiacal Light (ZL), and  
$I_{\rm DGL}$ is the Diffuse Galactic Light (DGL) surface brightness,
all to be determined in the direction of the BFM02 target field at $l = 206.6, 
b = -59.8$ deg.
Each one of the three components  is derived in BFM02 with a different method.
$I_{\rm tot}$ is measured above the atmosphere with the {\em Hubble Space Telescope} 
({\em HST}) using broad-band CCD photometry;
$I_{\rm ZL}$ is measured from the ground, with the 2.5-m du Pont telescope 
at the Las Campanas Observatory (LCO), 
using spectrophotometry; and 
$I_{\rm DGL}$ is estimated by using a model for the scattering of 
starlight by interstellar dust.
A very demanding task for the BFM02 method is set by the requirement that,
for each of the two telescopes with different properties and  different observing methods,  
the measured flux, $I_{\rm ZL}$ or $I_{\rm tot}$, has to be separately 
calibrated to the same scale. 
$I_{\rm EBL}$ is only a small fraction, a few per cent at most, of $I_{\rm tot}$ and
$I_{\rm ZL}$.
Therefore, the BFM02 method crucially depends on whether or not the very high
absolute accuracy, of $\lesssim$1 \%,  needed in the measurement, calibration, 
and scattered light corrections for
$I_{\rm ZL}$ and separately for $I_{\rm tot}$ is achieved. 

In this paper a critical discussion will be presented 
of the calibration and the atmospheric corrections applied in BFM02b 
to the ground based measurement of the Zodiacal Light. 
The assumptions for estimating $I_{\rm DGL}$ are critically reviewed.
In addition, I point out  a puzzling situation which emergerges from 
comparison between the results of BFM02 and 
of the widely cited previous presentations of the same observations.   

\section{Ground based measurement of the Zodiacal Light}

Two atmospheric corrections have to be applied:
(1) extinction, and (2) tropospheric scattered light.
The observed night sky brightness,  $I_{\rm obs}(\lambda,t,X)$, 
towards the target field is given by
\begin{eqnarray}
&&I_{\rm obs}(\lambda,t,X) =  \nonumber\\
&&I_{\rm ZL}(\lambda)e^{-\tau(\lambda)X} + [I_{\rm EBL}+ I_{\rm DGL}](\lambda)e^{-\tau(\lambda)X} + \nonumber\\
&&I_{\rm sca}(\lambda,t,X) + I_{\rm agl}(\lambda,t,X)
\end{eqnarray}
where $\lambda$ is the wavelength, $t$ the time of the observation, $X$ the airmass,
$\tau(\lambda)$ the atmospheric
extinction coefficient for unit airmass, 
$I_{\rm sca}$ the tropospheric scattered light,
and $I_{\rm agl}$ the airglow as observed from the ground (including atmospheric
attenuation and  scattered airglow). 
The BFM02b method of separating the ZL from the airglow component 
is based on
the assumptions that (1) the depths of the Fraunhofer lines in the spectrum of
the Zodiacal Light are identical to those in the solar spectrum, and 
(2) the airglow spectrum is uncorrelated with the solar Fraunhofer spectrum. 

The tropospheric scattered light, $I_{\rm sca}(\lambda,t,X)$, is the main obstacle
in conducting accurate {\em absolute} diffuse sky photometry from the ground. Unlike the 
photometry of stars or small extended sources, no differential ON/OFF measurements
are possible, and one must calculate the scattered light contribution coming
from all the light sources above 
the horizon. The scattered light components which contribute to the ``ZL-like'' (Fraunhofer 
spectrum) signal are due to the all-sky distributions of the ZL itself, the 
Integrated Starlight (ISL), and the DGL.
Each one of these components has 
the Rayleigh (R) and the Mie or aerosol (M) scattering part:
\begin{eqnarray}
&&I_{\rm sca}(\lambda,t,X)=I^{\rm R}_{\rm sca}({\rm ZL})+I^{\rm M}_{\rm sca}({\rm ZL})+  \nonumber\\ 
&&I^{\rm R}_{\rm sca}({\rm ISL})+I^{\rm M}_{\rm sca}({\rm ISL})+ I^{\rm R}_{\rm sca}({\rm DGL})+I^{\rm M}_{\rm sca}({\rm DGL}) \nonumber
\end{eqnarray}

\newpage

\section{Comparison between BFM02 and the previous results of Bernstein et al.}

The results of BFM02 have been previously presented in 
in  \citet[]{ber98,bea99,beb99,ber01}, and distributed in a preprint form 
for some time \citep[]{bfm99,bfm00}; in the following they are collectively referred
to as BFM98-01.
These results have been widely cited in the literature 
as the ``EBL standard reference values'', see e.g.  \citet[]{bar01, hau00, jim99,
lon01, mad00, pag01,pee01, poz01, pri99, ren01, wri01}. It is therefore  
important to point out the {following puzzling situation which results from the
comparison} between the BFM98-01 and BFM02 results:

In BFM98-01, the reduction method for 
the Zodiacal Light measurements differed in a fundamental way from that in BFM02b. 
Instead of Eq.(2), the following formula was applied:
\begin{eqnarray}
&&I_{\rm obs}(\lambda,t,X) =  \nonumber\\
&&I_{\rm ZL}(\lambda)e^{-\tau(\lambda)X} + [I_{\rm EBL}+ I_{\rm DGL}](\lambda)e^{-\tau(\lambda)X} + \nonumber\\ 
&&I_{\rm agl}(\lambda,t,X)
\end{eqnarray}
i.e. the atmospheric scattered light term $I_{\rm sca}(\lambda,t,X)$ 
was completely omitted without explanation, thus neglecting
this fundamental aspect of the diffuse night sky photometry.
Only in BFM02b has an  Appendix on scattered light model calculations
now been added. 
Using the results of these calculations a lower limit to  
$I_{\rm sca}$  at 4650 \AA\, can be estimated by adopting $X=1.1$ for the airmass, i.e. 
a value at the lower end of the BFM02b airmass range.
$I_{\rm sca}$ amounts to $\sim$18\% of $I_{\rm ZL}$.
It consists of the Rayleigh ($\sim$10\% of $I_{\rm ZL}$) and Mie ($\sim$4\% of  $I_{\rm ZL}$) components 
of scattered Zodiacal Light, 
as well as  of a scattered ISL component with a ``ZL-like'' spectrum ($\sim$4\% of $I_{\rm ZL}$) 
(see Sect. 4 and 5, and Figs.20, 21, and 30 of BFM02b). 
This value translates, for the airmass of 1.1,
to an outside-the-atmosphere value of $\sim$22\% of $I_{\rm ZL}$ which, for
the BFM02b value of  $I_{\rm ZL} = 109.4\times$\cgs at 4650 \AA\,, corresponds to 
$\sim$24$\times$\cgs.

Since not subtracted, $I_{\rm sca}$ was 
erroneously included into the Zodiacal Light, $I_{\rm ZL}$, in the BFM98-01 analysis.
Thus, the difference in the reduction methods 
necessarily should have resulted in a $\sim$22\% larger ZL value in BFM98-01 than in BFM02b.
This $\sim$22\% difference in $I_{\rm ZL}$ is
at least 7 times as large as the EBL value of BFM02 at 5500 and 8000 \AA. 
However, in BFM98-01  the EBL intensities at 3000, 5500, and 8000 \AA\,  are identical 
to the values in BFM02 of 4.0, 2.7, and 2.2$\times$\cgs\, to within 
$\leqq 0.1\times$\cgs.
It is unclear how this puzzling situation should be understood.

\section{Scattered light corrections}

In an Appendix of BFM02b model calculations for the atmospheric scattered light
are now presented. However, the nature of the problem dictates that it is 
hardly possible to achieve an absolute accuracy of $\lesssim$1$\times$\cgs\,
as required in the BFM02 method. Major problems are caused by e.g.
the varying properties of the atmospheric aerosols and ground reflectance, 
as well as by the insufficiently 
known intensity distributions of the main light sources, the ZL, ISL, and DGL over the sky. 

{\ The Moon, if above or up to a few degrees below the horizon, will give rise to a substantial
atmospheric scattered light component \citep[]{kri91}.
The observations at LCO, if carried out on 1995 November 27 and 29 as announced by 
BFM98-01 and BFM02b, 
would have included time slots when the Moon 
was above the horizon. In an erratum \citep[]{ber03} the authors have now removed this
problem by stating that their published dates were incorrect, the correct dates of their 
observing nights being 1995 November 24/25 and November 26/27.} 
 
\subsection{Aerosol scattering}

BFM02b have adopted for the albedo of the aerosol particles a value of $a =0.59$. 
This value, given in \citet[]{sta75}, was calculated for an ad hoc particle 
composition with a refractive index of $m = 1.5-0.1i$.
For a realistic aerosol composition, according
to  \citet[]{gar91} and  \citet[]{mcc78}, the albedo is $\sim$0.94.
{This is a representative value over the whole wavelength range, 3900 - 5100 \AA\,,
of the BFM02b Zodiacal Light measurement} and for a variety of different measured 
aerosol populations and conditions. { The 
aerosol scattering component, 
$I_{\rm sca}^{\rm M} = I_{\rm sca}^{\rm M}({\rm ZL})+I_{\rm sca}^{\rm M}({\rm ISL})$, 
has to be corrected for the effect of increased albedo 
by multiplying it with the albedo ratio 0.94/0.59.  
In order to obtain the contribution to the extraterrestrial  
$I_{\rm ZL}$ value one has to multiply  $I_{\rm sca}^{\rm M}$ by the extinction correction 
factor  $e^{\tau(\lambda) X}$.

{ According to BFM02b Appendix,  $I_{\rm sca}^{\rm M}(\rm ZL)$
is 3.6 - 6.1 \% of $I_{\rm ZL}$ 
at their minimum airmass of 1.02
and  5.2 - 8.1 \% of $I_{\rm ZL}$ at their maximum airmass of 1.82.
The lower value stands for  $\lambda = 4920$ \AA\,  and the higher one for
$\lambda = 3960$ \AA.  ({ Four selected wavelength slots at $\sim$3960, 4270,
4340, and 4920 \AA\, were used by BFM02b to derive their final $I_{\rm ZL}$ value 
at 4650 \AA.)} 

$I_{\rm sca}^{\rm M}(\rm ISL)$  is not explicitly given in BFM02b Appendix. 
It is only a small fraction of the total (Rayleigh + Mie) scattered ISL 
which they estimate to be  $\sim 12 - 24$ $\times$\cgs. A rough estimate is
obtained from} 
\begin{equation}
I_{\rm sca}^{\rm M}({\rm ISL}) \approx  a (1 - e^{-\tau_M X}) I_{\rm ISL},
\end{equation}
{ where $\tau_M$ is the aerosol extinction per airmass 
and  $I_{\rm ISL} \approx 40$ $\times$\cgs\, is the ISL intensity in the
direction of the target. For the
wavelength and airmass range, and the albedo $a = 0.59$, as adopted by BFM02b,
$I_{\rm sca}^{\rm M}(\rm ISL) \approx (0.9 - 2.4)$ $\times$\cgs. A third of $I_{\rm sca}^{\rm M}(\rm ISL)$ 
is estimated to have a solar type Fraunhofer spectrum and
thus adds  $\sim$0.3 - 0.8 \% of $I_{\rm ZL}$ to the scattered ZL value.} 
 
{ The average of $I_{\rm sca}^{\rm M} e^{\tau(\lambda) X}$
for the 16 different aimasses and four wavelength slots 
of the BFM02b measurement gives an  
outside-the-atmosphere value of  $\sim$7.6 \%  of  $I_{\rm ZL}$.
With the aerosol albedo   
of 0.94 instead of 0.59, this value should be scaled up to $\sim$12.1 \% of $I_{\rm ZL}$.
The difference of  $\sim$4.5 \%  of  $I_{\rm ZL}$ corresponds to 4.9$\times$\cgs\,
at 4650 \AA\, which, when not subtracted, will artificially increase the $I_{\rm ZL}$ 
value (see Table 1).}

\newpage
  
\subsection{Ground reflection}

BFM02b have neglected the reflection from the ground. 
{ Light reflected from the ground is scattered a second time by the
molecules in the atmosphere, back into the observer's line of sight.
Because of the strongly forward scattering phase function of the aerosols,
only the Rayleigh scattering by molecules is important here.} 
The influence of the ground reflection on the Rayleigh scattered light 
intensity can be estimated using the tables of \citet[]{ash54}.
The ground reflectance value 
obtained from the NASA MODIS/Terra Surface Reflectance database
\footnote{Available at http://modis-land.gsfc.nasa.gov/mod09/} is $\sim$8 \%. 
This value is for a 
100$\times$100 km$^2$ area centered at the Las Campanas Observatory, for the same 
season of the year 
(November-December 2000 and 2001) as the BFM02b observations, and for 
a wavelength band  (4590 - 4790 \AA\,, MODIS/Terra band 3) which closely matches the 
one used in BFM02b. According to the
 \citet[]{ash54} tables the Rayleigh scattered light intensity,
for the extinction range of BFM02b,
has to be increased by 7.2\% relative to the case of zero ground reflectance.

{ According to  BFM02b Appendix,  $I_{\rm sca}^{\rm R}(\rm ZL)$
is 8-18 \% of $I_{\rm ZL}$ at their minimum airmass of 1.02
and 15-32 \% of $I_{\rm ZL}$ at their maximum airmass of 1.82.
The lower value stands for  $\lambda = 4920$ \AA\,  and the higher one for
$\lambda = 3960$ \AA. The value of 
$I_{\rm sca}^{\rm R}(\rm ISL)$  is not explicitely given in BFM02b, but it can be
estimated using the total (Rayleigh + Mie)
scattered ISL which BFM02b estimate to be  $\sim 12 - 24$ $\times$\cgs. 
After subtracting  $I_{\rm sca}^{\rm M}$(\rm ISL) according to Sect. 4.1
and adding  $I_{\rm sca}^{\rm R}$(\rm DGL)according to Sect. 4.3
one obtains the range  $I_{\rm sca}^{\rm R}(\rm ISL)+ I_{\rm sca}^{\rm R}(\rm DGL)= (14-27)$ $\times$\cgs\,. 
 A third of it is estimated to have a solar type Fraunhofer spectrum and
thus adds $\sim$4.6 - 8.3 \% of  $I_{\rm ZL}$ to the scattered ZL value.}

{ The average of the ground reflection correction 
referred to outside-the-atmosphere, 
$\Delta I_{\rm sca}^{\rm R} e^{\tau(\lambda) X}$,
at the 16 different aimasses and four wavelength slots of the BFM02b measurement, 
gives a value of  $\sim$2.1 \%  
of $I_{\rm ZL}$. This corresponds to 2.3 $\times$\cgs\,  at 4650 \AA\, 
which, when not subtracted, will artificially increase the  $I_{\rm ZL}$ value 
(see Table 1).}
  
\subsection{Diffuse Galactic Light as source}

BFM02b have constructed an approximate
model for the intensity distribution and spectrum of the ISL. They find that their 
averaged intensities  are within 10 \% of the star count integrations of 
\citet[]{roa61}. However, the ISL is only a part of
the Galactic radiation field which contributes to the tropospheric
scattered light, with the other part being the DGL. 
The contribution of the DGL to the total Galactic radiation
field can be estimated using the tabulation of the ratios 
$I_{DGL}/I_{ISL}$. These data are based on the {\em Pioneer 10} measurements at 4400 \AA\,  
by \citet[]{tol81} and are reproduced in Table 39 of \citet[]{lei98}.  
{ The  $I_{DGL}/I_{ISL}$ ratio varies as a function of galactic
latitude, from $\sim$0.25-0.34 at low galactic latitudes to
 $\sim$0.12 at high latitudes. There is no pronounced dependence of this
ratio on the galactic longitude. To estimate the total contribution
of the DGL to the Galactic radiation field a weighted mean value
of the ratio over the galactic latitude range was formed.
The overall ratio is 0.24, i.e. the pure ISL sky brightness used
by BFM02b as the Galactic illumination source has to be increased by
24\%.} This 24\% correction should be applied to the Rayleigh scattering
component only. Because of strong forward scattering, the Mie (aerosol)
component couples to the DGL in the viewing direction at high galactic
latitudes where $I_{DGL}/I_{ISL} \approx 0.12$.

{According to BFM02b, the scattered (Rayleigh+Mie) ISL is 12 - 24$\times$\cgs\,
over their airmass range of 1.02 to 1.82. The Mie component
has been estimated in Sect. 4.1 to be $I_{\rm sca}^{\rm M}(\rm ISL) \approx (0.9-2.4)$$\times$\cgs\,.
Thus, the Rayleigh component is  $I_{\rm sca}^{\rm R}(\rm ISL) \approx (11-22)$$\times$\cgs.
The atmospheric scattered DGL 
component is 24\% of it, i.e. $I_{\rm sca}^{\rm R}(\rm DGL)\approx (2.6-5.2) \times$\cgs\,.}  
However, as discussed in BFM02b, it is not the total scattered 
flux which is important, but rather the strength of the Fraunhofer spectral features
which are in common with the Sun. According to BFM02b, the strengths of the
spectral features they use in their analysis are approximately 1.5 to 3.8 times
weaker in the DGL than in the ZL spectrum. { To obtain a reasonable lower-end 
estimate for the ``ZL-like'' component I use here a factor of 3.0, leading to
a range of (0.9-1.8)$\times$\cgs\,.}

{ The average of $I_{\rm sca}^{\rm R}(\rm DGL)e^{\tau(\lambda) X}$
at the 16 different aimasses and four wavelength slots
of the BFM02b measurement 
gives an  outside-the-atmosphere value of  $\sim$1.6$\times$\cgs\, at 4650 \AA\, 
which, when not subtracted, will artificially increase the  $I_{\rm ZL}$ value 
(see Table 1).}

\section{Calibration of the Zodiacal Light measurements: aperture correction}

{As stated in BFM02a,b the calibration of extended uniform surface brightness photometry
differs in an essential way from point source photometry. The following two 
standard issues are in common with point source and surface photometry:\\
(1) Point source calibration (including the atmospheric extinction effects); and
(2) Calibration of the fiducial standard star system.\\ 
However, in the calibration of extended uniform surface brightness photometry 
one requires knowledge of two additional aspects:\\
(3) The solid angle subtended by the spectrometer or photometer aperture or 
CCD detector pixel; and
(4) The aperture correction factor which accounts for the loss of flux of the standard star
outside the spectrometer or photometer aperture.

Because of scattering by micro-roughness and dust particle contamination on the optics,
and atmospheric small-angle scatting, the point spread function of a telescope
extends beyond the aperture size of 1-20 arcsec normally used in point source
photometry. The aperture correction factor, $T(A)$, is the fraction of the flux
from a point source that is contained within the aperture. The fraction $1 - T(A)$
of the point source flux is lost outside the aperture. In the measurement
of a uniform extended source the situation is different: the flux which is lost
from the solid angle defined by the focal plane aperture is compensated for
by the flux which is scattered and diffracted into the aperture from the sky
outside of the solid angle of the aperture. Therefore, the intensity 
in units of erg\,s$^{-1}$cm$^{-2}$sr$^{-1}$\AA$^{-1}$\, of an extended uniform 
source is given by 
\begin{equation}
I(\lambda) = \frac{S(\lambda) T(A)}{\Omega} C(\lambda),
\end{equation}
where  $C(\lambda)$ is the signal in instrumental units (DN s$^{-1}$),
$\Omega$  the solid angle of the aperture in steradians, and $S(\lambda)$ the sensitivity
function in units of erg s$^{-1}$ cm$^{-2}$ \AA$^{-1}$/(DN s$^{-1}$) determined from the standard 
star observations.

The aperture correction is an additional factor specific to the  
absolute surface photometry of extended uniform sources. 
It does not appear in point source photometry, nor in the photometry of small 
extended sources where the sky background can be measured next to the object.
The aperture correction factor is not ``automatically'' taken care of by simply using in the
standard star observations the same measuring arrangement as was used in
generating the standard star system.

The  accuracy of the EBL measurement of BFM02 crucially depends
on how consistently the surface brightnesses for the  
{\em HST} and du Pont telescope can be calibrated to the same scale
in spite of the different properties of the two telescopes. 
In practice, as also pointed out by BFM02, the majority of error in
calibrating a uniform surface brightness source comes from the accuracy with
which the large-angle PSF is determined.  
In fact, the two telescopes differ strongly
with respect to the aperture correction factor.} 

The surface (spectro)photometry of BFM02b at LCO was calibrated by standard
stars which were observed through a slit of 10.8 arcsec width. In order  
to compensate for the light lost outside of the slit, BFM02b measured the PSF
up to a radius of $\sim$60 arcsec. They found that the aperture correction factor
for a uniform-surface-brightness, aperture-filling source is 
$T = 0.963$. However, in their analysis,
BFM02b did not take into account that, in order to include 100\% of encircled energy
into the star image, the integration must be extended far beyond 
60 arcsec from the axis.  

A widely used compilation of data for the profile of a star image 
was presented by \citet[]{kin71}. Outside the central
disk of $\sim$10 arcsec radius, the image shows a more slowly declining 
halo or aureole which is well represented by an inverse-square law of intensity over 
a factor of 1000 in angular distance. 
Between 1 and 100 arcmin 
the aureole contains about 5\% and between 1 arcmin and 5 deg about 6\%
of the star's flux. 
Later work by e.g. 
\citet[]{cap83} (McDonald 0.9-m and 2-m telescopes), 
\citet[]{sur90} (Calar Alto 1.23-m),
\citet[]{rac96} (CTIO 4-m),
\citet[]{mac92} (KPNO 0.6-m Burrel Schmidt),
\citet[]{mid89} (La Palma ING 2.5-m), \citet[]{pic73} (Goethe Link 16-inch and McDonald 2-m), 
and \citet[]{uso91} (KPNO No.\,1 0.9-m) 
has demonstrated that very similar functional dependences (i.e. inverse square law), 
but in some cases substantially higher aureole energy fractions, 
are obtained for other telescopes equipped with photoelectric photometers, 
CCD cameras or spectrographs. For example,  \citet[]{cap83} find an integrated aureole 
contribution of $\sim$10\% between 30 arcec and 1.5 deg. Furthermore, in \citet[]{uso91}, 
the aureole stray radiation level is  $\sim$1.5 times higher than in \citet[]{kin71}. 
The probable reason for the aureole is
light scattering by the imperfections of the telescope optics, such as microripples 
and dust on the mirrors \citep[]{bec95,rod95}.

Although no stellar image aureole measurements were available in BFM02b or BFM98-01
for the du Pont telescope, there are good reasons to assume that its characteristics
are similar to the telescopes mentioned above. 
Including the aureole contribution to the stellar PSF, the aperture correction factor
of BFM02b should be decreased by 5 - 10\%, from $T = 0.963$ to 0.86 - 0.91.
The resulting outside-the-atmosphere ZL value should thus be reduced by 5 - 10\%, 
corresponding to 5.5 - 11$\times$\cgs\, at 4650 \AA\,(see Table 1). 

{ For their {\em HST} measurement BFM02a used the WFPC2 default calibration and adopted 
the standard aperture corrections as given in \citet[]{hol95}. For their filters used,
they applied an aperture correction factor of 90 \% in moving from a 0\farcs5 radius 
aperture to an ``infinite'' aperture (6\farcs0 radius).  Large
angle scattering measurements by \citet[]{kri94} at 20\arcsec- 60\arcsec have shown that 
the scattered light level in {\em HST}/WFPC2 is much below the aureole levels 
of the ground-based telescopes as listed above.  \citet[]{uso91} and \citet[]{sur90}
give at 60\arcsec a stray radiation value of 15.8 mag arcsec$^{-2}$ (normalisation to 
a $0^{\rm th}$ mag star), 
corresponding to 4.8$\times$10$^{-7}$ arcsec$^{-2}$ (normalisation to stellar flux = 1), 
while the {\em HST}/WFPC2 value is 7$\times$10$^{-8}$ arcsec$^{-2}$, i.e. a factor of $\sim$7 lower.
This is as expected since the {\em HST} optics are
known to have extremely small microroughness and are practically dust-free. Therefore, 
it is unlikely that any additional
large-angle aureole correction to the {\em HST} aperture correction factor is needed 
beyond the 90\% correction applied by BFM02a. In any case it  
would be much smaller than the correction for the ground-based du Pont telescope. 
Thus, when taking the difference  
$I_{\rm tot} - I_{\rm ZL}$,
the error in the aperture correction factor for $I_{\rm ZL}$ (du Pont telescope)
is not compensated for by a similar error in  $I_{\rm tot}$ ({\em HST}) .}

\begin{table*}[t]
\begin{center}
\caption{Corrections to be applied to the data analysis of BFM02: 
$\Delta I_{0}$ is the correction above-the-atmosphere. 
The corrections 1-4
are to be subtracted from the  $I_{\rm ZL}$ values, the DGL correction (item 5)
is to be added to the  $I_{\rm DGL}$ value as given in BFM02. The last three columns 
give the resulting 
corrections to be applied to the BFM02  $I_{\rm EBL}$ values at 3000, 5500, and 8000 \AA.
In lower part of the table the sum of the corrections 1 - 5,
the BFM02  $I_{\rm EBL}$ values at 3000, 5500, and 8000 \AA\,, 
and the  $I_{\rm EBL}$ estimates of this paper after applying the corrections are given.}

\scriptsize

\label{tbl-1}
\begin{tabular}{|llcc|cl|ccc|}
\tableline\tableline

 &Correction  & BFM02 & This paper&
 \multicolumn{1}{c}{$\Delta I_{0}$\tablenotemark{a}} & Applies to & 
& $\Delta I_{\rm EBL}$\tablenotemark{b}&   \\
 \multicolumn{2}{|c}{ } & &  & & & @3000 \AA &  @5500 \AA & @8000 \AA \\
\tableline
1.& Aerosol albedo            &0.59&0.94    & -4.9 & ZL @ 4650 \AA      & +1.5  & +4.7& +3.2  \\
2.& Ground reflectance        & neglected &8\%  & -2.3 &  ZL @ 4650 \AA & +0.7  & +2.2& +1.5  \\
3.& DGL as source of          & & &   &       &  & &  \\
  & atmospheric scattering    & neglected &included  & -1.6 &   ZL @ 4650 \AA &+0.5  & +1.5 & +1.1 \\
4.& ZL calibration,           & & &        &  &  & &  \\
  &aperture correction factor &0.963&0.86-0.91 & -5.5to-11 &   ZL @ 4650 \AA &+1.7-3.4  & +5.3-10.6 &+3.6-7.3  \\
5.& DGL modeling,             & & &   &      & & &    \\
  &line-of-sight $A_V$        &0$\fm$028&0$\fm$081$\pm$0$\fm$025   & +1.5to+1.9 & DGL @ all $\lambda$ & -1 & -1 & -1\\ 
\tableline
 &                            & &    &     &  &          &            &            \\
 & Sum of corrections 1 - 5   & &    &     &  &+3.4-5.1  & +12.7-18.0 & +8.4-12.1  \\
\tableline
 &                            & &    &     &  &          &            &            \\
 & $I_{\rm EBL}$  (BFM02)        &   &  &        &  & 4.0$\pm$2.5 &  2.7$\pm$1.4 & 2.2$\pm$1.0 \\
 & $I_{\rm EBL}$  (this paper) &  &  &        &  &7.4-9.1  & 15.4-20.7 &10.6-14.3  \\
\tableline
\end{tabular}
\tablenotetext{a}{in units of \cgs}
\tablenotetext{b}{Corrections in units of \cgs\, to be applied to the BFM02 $I_{\rm EBL}$ 
values}

\end{center}
\end{table*}

\section{Model estimates of the Diffuse Galactic Light}

The BFM02a estimation of the DGL intensity is based on the 100 $\mu$m surface
brightness,  $I_{100}$, as extracted from the IRAS Sky Survey Atlas (ISSA). This 
Atlas gives the very low value of 0.4 MJy sr$^{-1}$  for their target position.
The authors have apparently not paid attention to the fact that the
ISSA surface brightnesses cannot be utilised for {\em absolute} surface photometry;
see The Explanatory Supplement to the IRAS Sky Survey Atlas,  
\citet[]{whe94}, p. I-6. 
From their $I_{100}$ estimate BFM02a derived the values 
$N({\rm H}) = 0.47 \times 10^{20}$cm$^{-2}$ and $A_V = 0.028$ mag.
However, the value of $N$(H) can be directly extracted from the \citet[]{har97} Atlas,
carefully corrected for stray radiation effects. It gives for the 
BFM02 target  $N({\rm H}) = 1.78 \times 10^{20}$cm$^{-2}$ corresponding to
 $A_V = 0.106$ mag, i.e. 3.8 times
as large as the value adopted in BFM02a. Using the COBE/DIRBE surface photometry
to fix the zero point for the IRAS 100 $\mu$m data, \citet[]{sch98}
produced an all-sky Galactic  100 $\mu$m emission and optical extinction atlas. 
The values derived for the BFM02 target direction are   $I_{100}$ = 0.8 MJy sr$^{-1}$
and  $A_V = 0.054$ mag, i.e. twice as large as the value adopted by BFM02. 
From these two determinations, one obtains an average correction factor of 
$2.9\pm0.9$ by which the BFM02a 
dust column density and DGL intensity estimates have to be multiplied. The resulting
DGL intensities then become 2.9, 2.3, and 2.3 $\times$\cgs at 3000, 5500, 
and 8000 \AA\,, respectively. These are to be compared with the $I_{\rm DGL}$ values of
1.0, 0.8, and 0.8 $\times$\cgs as given in BFM02a at the three wavelengths.
Thus, the DGL intensity is substantially larger than estimated by BFM02a and
about equal to their EBL intensities.  

Because the DGL has a similar spectrum as the ZL, part of the DGL has been
included into the ZL measurement and therefore has been subtracted from $I_{\rm tot}$ 
together with the ZL. According to BFM02, this part is $\sim$35 \%. 
Therefore, the remaining DGL correction, to be 
subtracted from  $I_{\rm EBL}$ as given in BFM02a, amounts to $\sim1\times$\cgs.

\section{The resulting EBL estimates and their errors}

My analysis above has revealed several, in part serious problems in the BFM02b treatment 
of the ground-based Zodiacal Light observations.\\ 
(1) There is a major puzzle resulting from the comparison of the BFM02b and  
BFM98-01 treatment of atmospheric scattered light:  in spite of a 
large difference in the scattered light corrections applied, $I_{\rm sca} = 0$ in BFM98-01 vs. 
$I_{\rm sca} \gtrapprox 24\times$\cgs in BFM02b,
the derived EBL values are exactly the same to within $0.1\times$\cgs.\\
(2) The  systematic errors in calibration and atmospheric scattering, discussed in Sections 4.1-4.3 and 5, 
influence the derived ZL value 
in the same direction (see Table 1); the corrected  $I_{\rm ZL}$ value at 4650 \AA\, is  smaller than the 
estimate given in BFM02b by $\sim$14.3-19.8$\times$\cgs\,. This means that the EBL 
estimates of BFM02 have to be increased by 
$\sim$4.4-6.1, $\sim$13.7-19.0, and $\sim$9.4-13.1 $\times$\cgs\, at  3000, 5500, and 
8000 \AA\,, respectively.\\ 
(3) BFM02a have based their estimation of the DGL intensity on
too low a dust column density. The error caused by this is $\sim$1$\times$\cgs, and
is to be subtracted from the BFM02 $I_{EBL}$ values. This correction is small
compared with the scattered light and calibration corrections described above and does not suffice to
compensate for them (see Table 1);\\
(4) The corrected EBL estimates will be increased to  $\sim$7.4-9.1, $\sim$15.4-20.7, and $\sim$10.6-14.3 
$\times$\cgs\, at  3000, 5500, and 8000 \AA\,, respectively.
These values are 2 to 7 times as high as the original BFM02 estimates. 
Clearly, such high EBL values are in conflict with the upper limits of $\sim$4.5-9$\times$\cgs\
as derived by \citet[]{dub79}, \citet[]{tol83}, and \citet[]{mat90} at 4000-5100 \AA.\\
(5) BFM02 give for their mean EBL intensities the 1$\sigma$ combined statistical and systematic 
uncertainties of 
$\pm$2.5, $\pm$1.4, and $\pm$1.0$\times$\cgs\, at  3000, 5500, and 8000 \AA\,, respectively. 
However, the systematic effects discussed in Sect. 3 - 6
of this paper have not been adequately dealt with in BFM02. The corrections as listed
in Table 1 are to be considered as systematic errors which are not included in the BFM02 error
analysis. They can be seen to substantially exceed the BFM02 1$\sigma$ combined statistical
and systematic error estimates.

In summary, I have presented arguments indicating that systematic errors of the BFM02 measurement
have been underestimated. The claim of the paper of a ``detection of the EBL'' appears premature. 


}

\end{document}